# Efficient cryogenic nonlinear conversion processes in periodically-poled thin-film lithium niobate waveguides

YUJIE CHENG,[*,1,2] XIAOTING LI,[*,3] LANTIAN FENG,[1,2,7] HAOCHUAN LI,[3] WENZHAO SUN,[4,5] XINYU SONG,[1,2] YUYANG DING,[6] GUANGCAN GUO,[1,2] CHENG WANG,[3,8] AND XIFENG REN[1,2,9]

[1]*CAS Key Laboratory of Quantum Information, University of Science and Technology of China, Hefei 230026, China.*
[2]*CAS Synergetic Innovation Center of Quantum Information & Quantum Physics, University of Science and Technology of China, Hefei 230026, China.*
[3]*Department of Electrical Engineering, City University of Hong Kong, Hong Kong, China.*
[4]*City University of Hong Kong (Dongguan), Dongguan 523808, China.*
[5]*Center of Information and Communication Technology, City University of Hong Kong Shenzhen Research Institute, Shenzhen 518057, China.*
[6]*Hefei Guizhen Chip Technologies Co., Ltd., Hefei 230000, China.*
[7]*fenglt@ustc.edu.cn*
[8]*cwang257@cityu.edu.hk*
[9]*renxf@ustc.edu.cn*
*\*These authors contributed equally to this work.*

**Abstract:** Periodically poled thin-film lithium niobate (TFLN) waveguides, which enable efficient quadratic nonlinear processes, serve as crucial foundation for classical and quantum signal processing. To expand their application scope, we provide the first investigation of nonlinear conversion processes in periodically poled TFLN waveguides at cryogenic condition (7 K). Through systematic experimental characterization, we find that the periodically poled TFLN waveguide retains its high conversion efficiency at both cryogenic and room temperatures for both classical second-harmonic generation and quantum photon-pair generation processes. Particularly, the photon-pair source at cryogenic condition shows high brightness (~8 MHz/μW) and broad bandwidth (>100 THz). These results demonstrate the significant potential of TFLN wavelength conversion devices for cryogenic applications and foster future scalable quantum photonic systems.

## 1. Introduction

Thin-film lithium niobate (TFLN), which features wide transparency window, good acousto-optic, electro-optic and nonlinear properties, has been considered as an all-around platform to construct highly efficient and multi-functional photonic integrated circuits (PICs) [1–7]. To further expand their application scope, it is important for TFLN devices to move beyond room temperature, toward operation at cryogenic conditions. Low-temperature environments could minimize thermal noise and thus are essential for most quantum systems, including superconducting circuits [8], defect centers in diamond [9], quantum dots [10], and trapped ions [11]. The long envisioned fully integrated scalable quantum PIC also requires interfacing TFLN devices with quantum light sources and detectors that operate at cryogenic temperatures [12–15]. Toward this direction, many pioneering research works have been reported in recent years. For instance, the integration of electro-optic modulators with superconducting nanowire single-photon detectors (SNSPDs), which work at cryogenic temperature, enables fully on-chip manipulation and detection of photonic quantum states [16].

Some basic optical modules, such as integrated optical filter [17], have also been successfully verified at cryogenic temperatures.

Apart from passive and electro-optic devices, TFLN photonic devices that make use of its nonlinear optical properties for the generation of new frequencies and signal conversion across different frequencies also play crucial roles in classical and quantum PICs. Among various schemes, quasi-phase matching (QPM) based on periodically poled TFLN devices is the most commonly adopted since it allows accessing the highest quadratic optical nonlinearity and a large nonlinear overlap simultaneously [18–22]. Many optical nonlinear applications under ambient conditions have been efficiently implemented in photonic integrated TFLN circuits. These include cascaded third-harmonic generation [23], frequency doubling and parametric oscillation [24,25] and photon-pair source preparation [26–28]. Demonstrating these nonlinear applications under cryogenic operation conditions will further improve the scalability of TFLN based PICs and promote optical interfacing among different quantum systems. Although optical nonlinear processes at cryogenic temperatures have been investigated in titanium in-diffused LN waveguides [29,30], the weak optical confinement in these waveguides still suffer from limited nonlinear wavelength conversion efficiency and challenges in constructing large-scale PICs.

In this letter, we report the first characterization of the quadratic optical nonlinearity, including classical second-harmonic generation (SHG) and quantum spontaneous parametric down-conversion (SPDC), in chip-integrated periodically poled TFLN waveguides at cryogenic temperature down to 7 K, which is the lowest temperature of our setup. We measure the dependence of QPM working wavelength on temperature and observe that the nonlinear wavelength conversion efficiency of TFLN waveguides at 7 K is comparable to that obtained at room temperature. Additionally, we achieve broadband and bright correlated photon-pair generation at cryogenic temperatures. Our results advance the development of cryogenic nonlinear photonics in compact TFLN waveguides and will contribute to the realization of integrated scalable quantum information applications.

## 2. TEMPERATURE DEPENDENCE ANALYSIS

The TFLN waveguide we use is fabricated from a 5 mol% MgO-doped $x$-cut LN-on-insulator wafer, as shown in Fig. 1a. The ridge waveguide, with silica cladding, exhibits a total LN layer thickness of 600 nm, an etching depth of 250 nm, and a waveguide top width of ∼1300 nm. The waveguide possesses a trapezoidal cross-section with sidewall angle of $\theta = 45^\circ$. During the fabrication process, we first employ electrical beam lithography (EBL) and thermal evaporation of the metal layer, followed by a standard lift-off process to pattern the poling electrodes. Periodic domain inversion is achieved by applying a sequence of high-voltage poling pulses with a peak voltage of ∼480 V and a duration of ∼1 ms [31]. Next, we remove the poling electrodes using metal etchant and then define waveguide patterns in the poled regions with aligned EBL and an Ar$^+$ ion-based reactive ion etching (RIE) process. The tight optical confinement and significant overlap among interacting photons enable efficient nonlinear frequency conversion in our designed nanophotonic waveguide. To fully utilize the highest second-order nonlinear tensor component, $d_{33}$ ($\chi^{(2)}_{zzz}$), in $x$-cut waveguides, all interacting waves are chosen to be in TE-polarized modes. Here, we focus on first-order periodic poling with poling period of $\Lambda = 4.3$ µm, designed by the quasi-phase matching condition. The theoretical conversion efficiency is estimated to exceed 4000%W$^{-1}$cm$^{-2}$ [18].

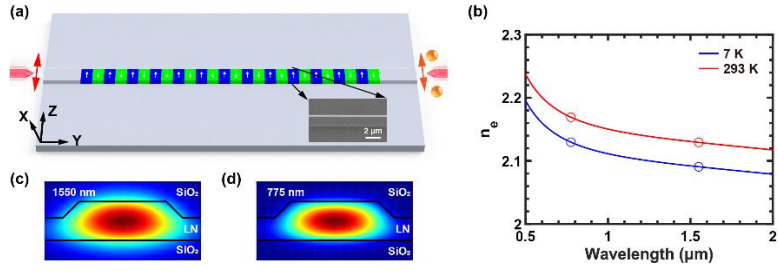

Fig. 1 Schematic configurations and simulation calculations. (a) Schematic illustration of the periodically poled TFLN waveguide. The total chip length is 1.1 cm, and the poled region has a length of 1 cm. end-fire coupling technique and lensed fibers are employed for photon input/output the waveguide. The on-chip optical loss at telecom wavelengths is estimated to be <1 dB. Inset: SEM image of the LN waveguide. (b) Calculated curves of the extraordinary refractive indices $n_e$ as a function of wavelength at 7 K and 293 K. We mark the values of $n_e$ at wavelengths of 775 nm and 1550 nm for both temperatures. (c) and (d) give the electric field distributions at cryogenic temperature (7 K) for wavelengths of 1550 nm and 775 nm, respectively.

To understand the effect of cryogenic condition on quadratic nonlinear processes, we first give a simple analysis on the phase matching condition at different temperatures. For the LN waveguide designed with the *x*-cut and *y*-propagation configuration, the effective indices of the guided transverse electric (TE) modes are mainly tied to the extraordinary refractive indices (that is, $n_e$) of the material. The variation of $n_e$ with temperature and wavelength is described by the Sellmeier equation [32]:

$$n_e^2 = a_1 + b_1 f + \frac{a_2 + b_2 f}{\lambda^2 - (a_3 + b_3 f)^2} + \frac{a_4 + b_4 f}{\lambda^2 - a_5^2} - a_6 \lambda^2, \qquad (1)$$

where the wavelength $\lambda$ has the unit of μm and the temperature dependent parameter *f* is given by:

$$f = (T - T_0)(T + T_0 + 2 \times 273.16). \qquad (2)$$

Here, $T_0 = 24.5°$ C and the temperature *T* is expressed in degree Celsius. The Sellmeier coefficients are: $a_1 = 5.756$, $a_2 = 0.0983$, $a_3 = 0.202$, $a_4 = 189.32$, $a_5 = 12.52$, $a_6 = 1.32 \times 10^{-2}$, $b_1 = 2.86 \times 10^{-6}$, $b_2 = 4.7 \times 10^{-8}$, $b_3 = 6.113 \times 10^{-8}$, and $b_4 = 1.516 \times 10^{-4}$. Using Sellmeier equation, we calculate the refractive index of LN at different temperatures and wavelengths of 775 nm and 1550 nm, as shown in Fig. 1b. Based on these values, we further obtain effective mode indices of $TE_0$ modes at various wavelengths and temperatures. Figures 1c and 1d show the field distributions at cryogenic temperature (7 K) for wavelengths of 1550 nm and 775 nm, respectively.

QPM is realized by periodic domain inversion of the LN waveguide, and it can be expressed in the context of SHG through the following equation:

$$\Delta k = k_{\text{SHG}} - k_{p,1} - k_{p,2} - \frac{2\pi}{\Lambda} = 0. \qquad (3)$$

In this equation, $\Delta k$ denotes the discrepancy between the propagation constants *k* of the interacting optical modes, $k_{\text{SHG}}$, $k_{p,1}$ and $k_{p,2}$ represent the wave vectors of second-harmonic photons and two pump photons, and $\Lambda$ is the poling period. Consider the influence of temperature, the formula is transformed to:

$$\frac{2\pi n_{\text{SHG}}(T, \lambda_{\text{SHG}})}{\lambda_{\text{SHG}}} - \frac{2\pi n_{p,1}(T, \lambda_{p,1})}{\lambda_{p,1}} - \frac{2\pi n_{p,2}(T, \lambda_{p,2})}{\lambda_{p,2}} - \frac{2\pi}{\Lambda(T)} = 0. \qquad (4)$$

Here, we plug $k = 2\pi n/\lambda$ into the formula. In SHG case, two pump photons are identical, hence $\lambda_p = \lambda_{p,1} = \lambda_{p,2} = 2\lambda_{\text{SHG}}$ and $n_p(T, \lambda_p) = n_{p,1}(T, \lambda_{p,1}) = n_{p,2}(T, \lambda_{p,2})$. Thus the equation is reduced to the following simplified form:

$$\Lambda(T) \cdot \left(n_{\text{SHG}}(T, \lambda_{\text{SHG}}) - n_p(T, \lambda_p)\right) = \lambda_{\text{SHG}}(T). \qquad (5)$$

Therefore,

$$\frac{\lambda_{SHG}(7K)}{\lambda_{SHG}(293K)} = \frac{\Lambda(7K)}{\Lambda(293K)} \times \frac{n_p(7K)-n_{SHG}(7K)}{n_p(293K)-n_{SHG}(293K)} \qquad (6)$$

Based on Eq. (6), we observe that the temperature causes the QPM working wavelength shift by affecting the mode effective refractive indices and the poling period $\Lambda$. Due to the extremely tiny thermal expansion coefficients of LN, which is on the order of $10^{-6}$/K [33], the variation of the cross-section and poling period of the waveguide at different temperatures can be considered insignificant. With $\Lambda$ (7 K) = $\Lambda$ (293 K), we get $\lambda_{SHG}$ (7 K)/$\lambda_{SHG}$ (293 K) ≈ 0.982. Assuming the phase matching wavelength of SHG at room temperature is 775 nm, it will be decreased to 761 nm at the temperature 7 K.

Next, we analyze temperature dependence of the normalized conversion efficiency $\eta$ with the following formula [18]:

$$\eta = \frac{2\omega^2 d_{eff}^2}{n_p^2 n_{SHG} \epsilon_0 c^3} \cdot \frac{A_{2\omega}}{A_\omega^2}, \qquad (7)$$

Where $\epsilon_0$ represents the vacuum permittivity, $c$ is the speed of light in vacuum, $d_{eff} = \frac{2}{\pi} d_{33}$ is the effective nonlinear coefficient, and $A_\omega$ and $A_{2\omega}$ represent the effective mode areas at wavelengths pump $\lambda_p$ and second harmonics $\lambda_{SHG}$, respectively. Based on Ref. [34], we assume that the highest nonlinear tensor component of LN $d_{33}$ is temperature-independent. We estimate that the conversion efficiency increases by only 3.8% at cryogenic temperatures compared to the result at room temperature, which is mainly due to the variation in mode area and the QPM wavelength at different temperatures.

## 3. CRYOGENIC SECOND HARMONIC GENERATION

The experimental setup to test cryogenic SHG of the periodically poled TFLN waveguide is shown in Fig. 2a. The nanophotonic chip and two lensed fibers for laser input to and output from the chip are placed in a cryostation chamber, which can control chip temperature from 7 K to room temperature with the precision of 20 mK. The lensed fibers, with a mode field diameter of 2.5 µm, are placed on two three-dimensional translational stage inside the chamber for coupling efficiency optimization. In the experiment, end-fire coupling technique is adopted and fiber-to-chip coupling loss is estimated to be 7 dB per facet at telecom wavelengths and 8 dB per facet at near-visible wavelengths. The collisions between the fibers and the chip could damage the chip, especially under cryogenic conditions where vibrations are more obvious. We slightly expand the coupling distance between the fibers and the chip to prevent potential collisions, although the coupling loss increased. A tunable telecom-band continuous wave (CW) laser is used as the pump. Before coupled into the chip, the polarization of the pump light is adjusted by a fiber polarization controller (FPC) to ensure TE-polarized mode propagation in the waveguide. After passing through the waveguide, the generated SH light is collected and guided out of the chamber by the lensed fiber and sent to a near-infrared (NIR) spectrograph for further analysis.

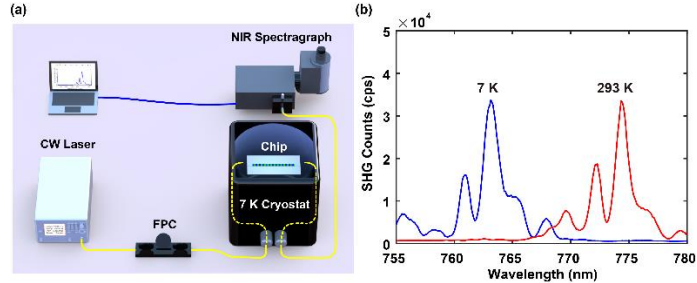

Fig. 2 Experiment setup and results for cryogenic SHG. (a) The experimental setup with the TFLN waveguide placed in the cryostation chamber. One tunable continuous wave (CW) laser is used as the pump, and its polarization is manipulated by one fiber polarization controller (FPC). The generated SH light is sent to a near-infrared (NIR) spectrograph for further analysis. (b) The measured SHG spectral responses of the periodically poled TFLN waveguide at temperatures of 7 K and 293 K.

The SH spectra at cryogenic and room temperature are shown in Fig. 2b. During the measurement, the pump power is kept at 20 µW (before coupling to the chip) and the spectrograph records SH single-photon counts at pump wavelength from 1505 nm to 1565 nm. Multiple tests have been conducted to obtain peak counts at each wavelength. The peak values are then connected to form a spectral curve. As shown in Fig. 2b, cryogenic temperature leads to shorter QPM wavelength, and the spectral shape and SH intensity are not significantly affected by temperature, as expected from our theoretical analysis above. Note that the measured SHG spectral response does not exhibit a theoretically sinc-function profile, which is attributed to possible poling non-uniformity and inhomogeneities of the film thickness [35]. The SH peak wavelength at room temperature (293 K) is 774.33 nm, whereas at the cryogenic temperature (7 K), it is shifted to 763.13 nm. The measured QPM wavelength blue-shift aligns well with the theoretical prediction given in Section 2. A more precise theoretical analysis can be achieved by introducing the effect of temperature on the refractive index of $SiO_2$ cladding [36,37].

## 4. CRYOGENIC PHOTON-PAIR GENERATION

Quantum light source is an indispensable module for quantum photonic information processing. Many previous reports have proved that periodically poled TFLN waveguides, which yield high quality correlated photon pairs, could provide high-performance solution for quantum light sources [26–28]. Therefore, in this paper, we further explore the possibility of realizing the SPDC nonlinear process in LN waveguides and thus achieving efficient quantum light sources at cryogenic temperatures. The experimental setup for testing cryogenic photon-pair generation is shown in Fig. 3a. Based on the cryogenic SH response measured in Fig. 2b, a CW laser centered at 763.13 nm is used as the pump light, and the input power is set as 1.34 µW (before coupling to the chip). The low input power minimizes the thermal effect on the material, making local variations in the refractive index across different positions less likely. The uniformity of the refractive index at both room temperature and cryogenic temperature contributes to the consistency of the effective length under these conditions. After passing through the waveguide, the generated correlated photon pairs are collected and guided out of the chamber by the lensed fiber. The photons then pass through cascaded filters to remove the pump light and are split equally into two paths by a 50/50 fiber beam splitter (BS). Two SNSPDs with dark counting rate of 100 Hz and detection efficiency of 80% are used to convert single-photon signals into electrical signals. Finally, the acquired electrical signals are collected and analyzed by a time-correlated single-photon counting system.

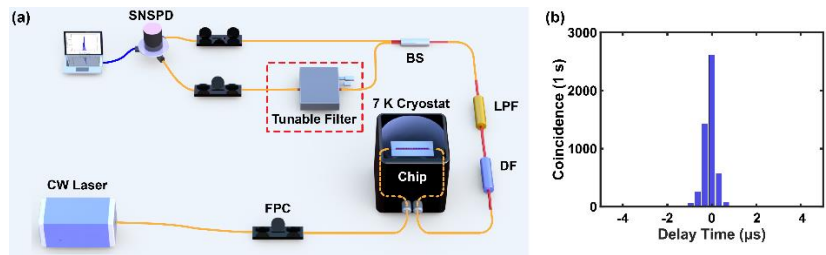

Fig. 3 Experiment setup and results for cryogenic photon-pair generation. (a) The pump laser is centered at 763.13 nm, which is selected based on the cryogenic SH response result. The generated photon pairs in the TFLN waveguide at 7 K first pass through cascaded filters to remove the pump light, and then are split equally into two paths by one 50/50 fiber beam splitter (BS) for coincidence measurement. The cascaded filters include one long-pass dichroic mirror (DM, transmission wavelength >950 nm) and one long-pass filter (LPF, cut-off wavelength 1400 nm). To measure the frequency bandwidth of the cryogenic quantum photonic source, one filter

with tunable center wavelength is additionally added to one output of fiber BS (dotted line framed), and specific wavelength is selected for coincidence measurement. SNSPD: superconducting nanowire single-photon detectors. (b) The measured temporal correlation coincidence of the cryogenic quantum photonic source.

With the coincidence time window set as 320 ps and integration time as 1 s, we record the temporal correlation coincidence between the two paths of photons, and the result is shown in Fig. 3b. The result is based on the averaged coincidence value obtained from two independent measurements. Despite the on-chip pump power being only a few hundred nanowatts, we have detected thousands of coincidence counts per second. Additionally, due to the existence of beam splitter, the actual photon-pair brightness is twice the number of photons detected. Accounting for coupling loss, filtering loss and detector efficiency, we estimate that the source brightness is about 8 MHz/µW.

Next, we consider the influence of pump power on the properties of cryogenic SPDC source. By varying the input pump power, we measure the two-photon coincidence and coincidence-to-accidental ratio (CAR) values as functions of the pump power with temperature kept at 7 K. The data are obtained with an integration time of 10 s, while the coincidence time window is maintained as 320 ps. Since two-photon coincidence values have peaks on multiple time channels (Fig. 3b), we record only the highest peak values (the same method is applied in the following experiments). The results are shown in Fig. 4a and 4b. The two-photon coincidence values show a linear relationship against the pump power, and the fitted curve of CAR results is obtained using the method introduced in Ref. [38]. The differences between the data and the fitted curves are mainly from the vibration of the test platform.

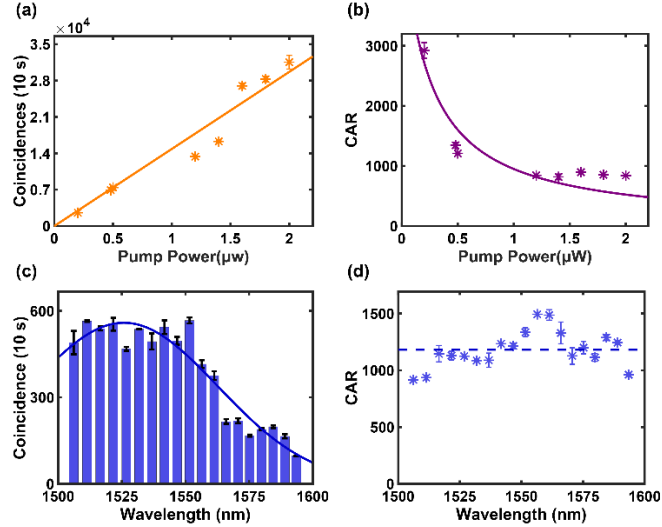

Fig. 4 Pump power dependence and multi-frequency channel characteristics of the SPDC photon-pair source at 7 K. (a-b) Two-photon coincidence counts (a) and CAR values (b) as functions of input pump power. (c) Two-photon coincidence at different center wavelengths of the tunable filter, which are normalized according to the filter transmittance and bandwidth. A Gaussian fitting curve is added for eye guidance, and the fitted 3 dB bandwidth is over 100 THz. (d) CAR values for different frequency channels. The dotted line indicates the average value.

In the SPDC process, one pump photon is converted to one time-correlated photon pair. Due to frequency dispersion compensation in TFLN waveguides, Eq. (3) is satisfied over a wide frequency band, enabling broadband quantum light sources. This shows significant advantages in multi-wavelength channel quantum information processing [39]. To investigate the emission spectrum of the cryogenic SPDC quantum photonic source in the TFLN waveguide, a tunable filter (dotted line framed in Fig. 3(a)) is added to one path of the fiber BS to select specific wavelength channels for further measurement. From 1500 nm to 1600 nm, we select 19 frequency channels for testing, and the measured channel bandwidth and central wavelength are provided in Supplement 1. The relationship between two-photon coincidences and filtering wavelength is shown in Fig. 4c. Although the obtained emission spectrum is not symmetrical due to the filter's tunable range limitation, we can still infer from the results that the radiation bandwidth is over 100 THz. CARs of different frequency channels are shown in Fig. 4d. Most values are over 1000 and average value is 1182 (the dotted line).

## 5. CONCLUSION

Our work provides a comprehensive analysis on second order nonlinear conversion processes, including SHG and SPDC, in periodically poled TFLN waveguides at cryogenic temperature (7 K). According to the experimental results, aside from the QPM wavelength shift, the cryogenic environment does not significantly impact the nonlinear conversion efficiency. The photon-pair source we prepared through the periodically poled TFLN waveguide shows high quality and broad bandwidth (>100 THz), showcasing its great potential in integration with other TFLN quantum devices that can work under cryogenic operation conditions, such as optical filters [17], optical modulators, and detectors [16]. Beyond photonic quantum applications, our work could also facilitate the development of other cryogenic nonlinear applications such as all-optical modulation and frequency comb generation with TFLN photonic integrated circuits.

**Supporting Information**
Supporting Information is available from the Wiley Online Library or from the author.

**Acknowledgments**
This work was supported by National Key Research and Development Program of China (2022YFA1204704), the National Science Foundation of China (NSFC) (62061160487, 62275240, T2325022, U23A2074, 62321166651), the CAS Project for Young Scientists in Basic Research (No.253 YSBR-049), Key Research and Development Program of Anhui Province (2022b1302007), Innovation Program for Quantum Science and Technology (No. 2021ZD0301500), the Fundamental Research Funds for the Central Universities, and the Research Grants Council, University Grants Committee (N_CityU113/20). This work was partially carried out at the USTC Center for Micro and Nanoscale Research and Fabrication.

**Conflict of Interest**
The authors declare no conflict of interest.

**Data Availability Statement**
Data underlying the results presented in this paper are not publicly available at this time but may be obtained from the authors upon reasonable request.

**Keywords**
thin-film lithium niobate waveguide, cryogenic, nonlinear conversion process, photon-pair source